\documentclass[sigconf]{acmart}
\pdfoutput=1
\DeclareMathDelimiter{(}{\mathopen} {operators}{"28}{largesymbols}{"00}
\DeclareMathDelimiter{)}{\mathclose}{operators}{"29}{largesymbols}{"01}
\onecolumn

\usepackage{subcaption}
\usepackage{graphicx}
\usepackage{hyperref}
\usepackage{setspace}
\usepackage{wrapfig}
\usepackage{amsmath}
\usepackage{times}
\usepackage{verbatim}
\usepackage{graphicx}
\usepackage{color}
\usepackage{url}
\usepackage{listings}
\usepackage{csquotes}
\usepackage[normalem]{ulem}
\usepackage{algorithm}

\usepackage[noend]{algpseudocode}
\setcopyright{none}
\renewcommand\footnotetextcopyrightpermission[1]{}

\algblock{Input}{EndInput}
\algblock{Output}{EndOutput}

\usepackage[normalem]{ulem}
\usepackage{amssymb}
\usepackage{xparse}
\usepackage{mathtools}

\makeatletter
\def\mdseries@tt{m}
\makeatother

\usepackage{multirow}
\usepackage{datapie}
\usepackage{adjustbox}
\def\BibTeX{{\rm B\kern-.05em{\sc i\kern-.025em b}\kern-.08emT\kern-.1667em\lower.7ex\hbox{E}\kern-.125emX}}

\newcommand{\lingming}[1]{{\color{red}{Lingming:[#1]}}}

\usepackage{xcolor}

\newcommand{\commit}{\textit{Revision no.}}
\newcommand{\simulee}[1]{\textit{Simulee}}
\newcommand{\gklee}[1]{\textit{GKLEE}}
\newcommand{\curd}[1]{\textit{CURD}}
\newcommand{\barracuda}[1]{\textit{BARRACUDA}}
\newcommand{\eraser}[1]{\textit{Eraser}}
\newcommand{\llvm}[1]{\textit{llvm}}
\newcommand{\codeIn}[1]{\texttt{#1}}
\newcommand{\remove}[1]{}
\newcommand{\git}[1]{\textit{GitHub}}
\newcommand{\junit}[1]{\textit{JUnit}}
\newcommand{\Always}[1]{\textit{Test failure}}
\newcommand{\always}[1]{\textit{test failure}}
\newcommand{\Sometime}[1]{\textit{Flaky test}}
\newcommand{\sometime}[1]{\textit{flaky test}}
\newcommand{\Crash}[1]{\textit{Crash}}
\newcommand{\crash}[1]{\textit{crash}}
\newcommand{\Perform}[1]{\textit{Inferior performance}}
\newcommand{\perform}[1]{\textit{inferior performance}}
\newcommand{\Improper}[1]{\textit{Improper resource management}}
\newcommand{\improper}[1]{\textit{improper resource management}}
\newcommand{\Non}[1]{\textit{Non-optimal implementation}}
\newcommand{\non}[1]{\textit{non-optimal implementation}}
\newcommand{\cross}[1]{\textit{generic error}}
\newcommand{\Cross}[1]{\textit{Generic error}}
\newcommand{\Port}[1]{\textit{Poor portability}}
\newcommand{\Sync}[1]{\textit{Improper synchronization}}

\newcommand{\port}[1]{\textit{poor portability}}

\newcommand{\sync}{improper synchronization}

\newcommand{\torder}[1]{\textit{Torder List}}
\newcommand{\distance}{0pt}

\newcommand{\targetFunction}{$F(dimensions, arguments)$}
\newcommand{\heuristic}{$f(i)$}
\newcommand{\first}{\textit{primary score}}
\newcommand{\second}{\textit{secondary score}}
\newcommand{\model}{\textit{Memory Model}}
\newcommand{\tuple}{\textit{Unit Tuple}}
\newcommand{\unit}{\textit{Memory Unit}}

\newcommand{\totalnew}{26}
\newcommand{\confirmnew}{10}
\newcommand{\totalproject}{9}
\newcommand{\newproject}{4}

\newcommand{\vo}[1]{\textit{visit\_order}}
\newcommand{\ti}[1]{\textit{thread\_id}}
\newcommand{\ac}[1]{\textit{action}}

\usepackage{caption}
\DTLloaddb{fruit}{test.csv}

\begin{document}

\title{Characterizing and Detecting CUDA Program Bugs}

\author{Mingyuan Wu}
\affiliation{%
  \institution{Department of Computer Science and Engineering, Southern University of Science and Technology}
  \city{Shenzhen}
  \state{Guangdong}
  \country{China}
}
\email{11849319@mail.sustech.edu.cn}

\author{Husheng Zhou}
\affiliation{%
  \institution{Department of Computer Science, The University of Texas at Dallas}
  \city{Richardson}
  \state{TX}
  \country{USA}}
  \email{husheng.zhou@utdallas.edu}

\author{Lingming Zhang}
\affiliation{%
 \institution{Department of Computer Science, The University of Texas at Dallas}
  \city{Richardson}
  \state{TX}
 \country{USA}}
\email{lingming.zhang@utdallas.edu}

\author{Cong Liu}
\affiliation{%
  \institution{Department of Computer Science, The University of Texas at Dallas}
 \city{Richardson}
  \state{TX}
  \country{USA}}
\email{cong@utdallas.edu}

\author{Yuqun Zhang}
\authornote{Corresponding author.}
\affiliation{%
  \institution{Department of Computer Science and Engineering, Southern University of Science and Technology}
  \city{Shenzhen}
  \state{Guangdong}
  \country{China}
}
\email{zhangyq@sustech.edu.cn}


\begin{abstract}
While CUDA has become a major parallel computing platform and programming model for general-purpose GPU computing, CUDA-induced bug patterns have not yet been well explored. In this paper, we conduct the first empirical study to reveal important categories of CUDA program bug patterns based on 319 bugs identified within 5 popular CUDA projects in \git{}. Our findings demonstrate that CUDA-specific characteristics may cause program bugs such as synchronization bugs that are rather difficult to detect. To efficiently detect such synchronization bugs, we establish the first lightweight general CUDA bug detection framework, namely \simulee{}, to simulate CUDA program execution by interpreting the corresponding \llvm{} bytecode and collecting the memory-access information to automatically detect CUDA synchronization bugs. To evaluate the effectiveness and efficiency of \simulee{}, we conduct a set of experiments and the experimental results suggest that \simulee{} can detect 20 out of the 27 studied synchronization bugs and successfully detects \totalnew{} previously unknown synchronization bugs\remove{for \totalproject{} 
projects (5 studied and \newproject{} new projects)}, \confirmnew{} of which have been confirmed by the developers.  
\end{abstract}
 \acmConference[ESEC/FSE 2019]
 {Paper submit to ESEC/FSE 2019}
 {26--30 August, 2019}
\maketitle
 
\section{Introduction} 
CUDA~\cite{cudawiki} is a major parallel computing platform and programming model that allows software developers to leverage general-purpose GPU (GPGPU) computing~\cite{gpgpuwiki}.  
CUDA is advanced in simplifying I/O streams to memories and dividing computations into sub-computations since it parallelizes programs in terms of grids and blocks. 
In addition, CUDA enables more flexible cache management that speeds up the floating point computation of CPUs. CUDA is thus considered rather powerful for accelerating deep-neural-network-related applications where the relevant matrix computations can be efficiently loaded.    
 
Due to the essential differences between GPUs and CPUs, traditional bug detection approaches for CPUs render inapplicable for GPUs. In particular, since GPU programs use barriers rather than locks for synchronization and enable simple happens-before relations, the traditional lockset-based ~\cite{lockset1, lockset2} and happens-before-based bug detection approaches \cite{Dinning:1990:ECM:99163.99165,happen2} for CPUs become obsolete in detecting the parallel-computing-related bugs for GPUs. On the other hand, it is argued that the lack of GPU parallel programming experience and the unawareness of implicit assumptions from the third-party kernel functions of developers are major reasons to cause GPU parallel-computing-related bugs. For instance, a developer might launch kernel functions with 512 threads when she is not aware that the optimal maximum number of threads in one block is only 256 \cite{Grace}. However, although CUDA programing has been dominating the popular deep-neural-networks-related applications, the studies on its parallel-computation-related bug patterns are rather limited \cite{icpp2012}. Therefore, a full scan of CUDA bug patterns could help developers understand the bug patterns to improve programming efficiency, and help researchers get enlightened for future research.

In this paper, we conduct a comprehensive empirical study on real-world CUDA bug patterns, based on 319 bugs collected from five popular CUDA projects with a total of 15314 commits and 1.1 million LOC (by Jan, 2019) 
in \git{} according to a set of policies that emphasize the importance and impact of these projects. 
Through the study, we build better understandings of the CUDA program bug patterns. In particular, we depict these collected CUDA program bugs by two dimensions: runtime stage and root cause. 
In our study, we identify three runtime stages and five root causes and obtain the following findings: 
(1) the majority of the kernel function bugs are not SIMD-specific only and can take place in other platforms, and thus can be detected by traditional CPU-based approaches;
(2) the majority of the memory-related bugs can also be solved by traditional approaches;
and (3) detecting synchronization bugs is important, challenging, and out of the scope of traditional approaches.

Inspired by these findings, we further develop a systematic lightweight bug detection framework, namely \simulee{}~\cite{simulee}, 
to detect the synchronization bugs for CUDA kernel functions. Though existing techniques, e.g., \gklee{}~\cite{gklee}, CURD~\cite{curd}, have been developed to detect CUDA program bugs, they are mostly either based on expensive static/dynamic analysis that results in non-negligible overhead, or fail to generate effective test cases to detect different types of synchronization bugs. \simulee{}, on the other hand, 
generates a \model{} that depicts the information regarding thread-wise memory access including thread id, visit order, and action. Accordingly, \simulee{} is launched by building a virtual machine that takes the \llvm{} bytecode of CUDA kernel functions for initializing the running environmental setups including arguments, dimensions, and global memory if necessary based on the \model{}. Next, \simulee{} applies Evolutionary Programming~\cite{evolutionaryprogramming} to approach error-inducing inputs for executing \llvm{} bytecode of CUDA programs and 
collects the corresponding memory-access information. 
At last, by combining CUDA specifics, such collected information can be analyzed to find whether they lead to synchronization bugs. 

Unlike other CUDA synchronization bug detection approaches that are mostly not fully automated~\cite{curd}~\cite{racecheckor}~\cite{barracuda} or limited in detecting certain bug types, e.g., data race \cite{Grace}~\cite{GMrace}, \simulee{} can detect multiple bug types including data race, redundant barrier function, and barrier divergence fully automatically. Moreover, \simulee{} benefits from only simulating runtime CUDA programs without incurring overhead for extra processing (e.g., searching), such that it is more efficient than the static/dynamic-analysis-based approaches~\cite{gklee, cudasmt} that usually demand large search space.

To evaluate the effectiveness and efficiency of \simulee{} on detecting synchronization bugs, we conduct a set of experiments based on a total of \totalproject{} projects, including the 5 projects for the empirical study and additional \newproject{} projects with in total 2113 commits and 122K LOC. 
The experimental results suggest that \simulee{} can successfully detect 20 of 27 synchronization bugs derived from the empirical study. It can further detect \totalnew{} previously unknown bugs of all the \totalproject{} projects, \confirmnew{} of which have already been confirmed by the corresponding developers. 
Moreover, the experimental results also demonstrate that \simulee{} can be much more effective than state-of-the-art \gklee{} in detecting synchronization bugs, e.g., none of the \confirmnew{} confirmed bugs can be detected by \gklee{}. 
In summary, our paper makes the following contributions:
\begin{itemize}
\item To the best of our knowledge, we conduct the first extensive study on the overall CUDA program bugs. Our findings can help understand the characteristics about CUDA bugs and guide the future relevant research . 
\item To the best of our knowledge, we develop the first lightweight, fully automated, and general-purpose detection framework for CUDA synchronization bugs, namely \simulee{}, that can automatically detect a wide range of synchronization bugs in CUDA programs which are hard to be captured manually. 
\item We evaluate \simulee{} under multiple experimental setups. The results suggest that \simulee{} is able to detect most of the synchronization bugs in the studied projects. In addition, it detected \totalnew{} new bugs of all the subject projects and outperforms state of the art. 
\end{itemize}

\remove{The rest of the paper is organized as follows. Section ~\ref{sec_bak} introduces the background of CUDA programs. Section ~\ref{sec_study} presents our empirical study. Section ~\ref{motivation} introduces the motivation of \simulee{}. Section ~\ref{sec_gunit} introduces \simulee{} and its effectiveness and efficiency to detect CUDA synchronization bugs. Section~\ref{sec_validity} and~\ref{sec_related} introduce the threats to validity and related work respectively. Section~\ref{sec_con} concludes this paper.}
\section{Background}
\label{sec_bak}
In this section, we give an overview on CUDA, the CUDA parallel computing mechanism, and typical CUDA synchronization bugs.

\textbf{CUDA Overview.} CUDA is a parallel computing platform and programming model, which enables developers to use GPU hardware for general-purpose computing. CUDA is composed of a runtime library and an extended version of C/C++. In particular, CUDA programs are executed on GPU cores, namely ``device'', while they need to be allocated with resources on CPUs, namely ``host'', prior to execution. As a result, developers need to retrieve allocated resources such as global memory after CUDA program execution. To conclude, a complete CUDA program contains 3 runtime stages: host resource preparation, kernel function execution, and host resource retrieve.

\textbf{CUDA Parallel Computing Mechanism.} Kernel function refers to the part of CUDA programs 
that runs on the device side. Specifically, thread is the kernel function's basic execution unit. In the \emph{physical} level, 32 threads are bundled as a thread warp wherein all the threads execute the same statement at any time except undergoing a branch divergence, while in the \emph{logic} level, one or more threads are contained in a block, and one or more blocks are contained in a grid.

 \begin{figure}[!t]
 	\centering
 	\includegraphics[width=0.3\textwidth]{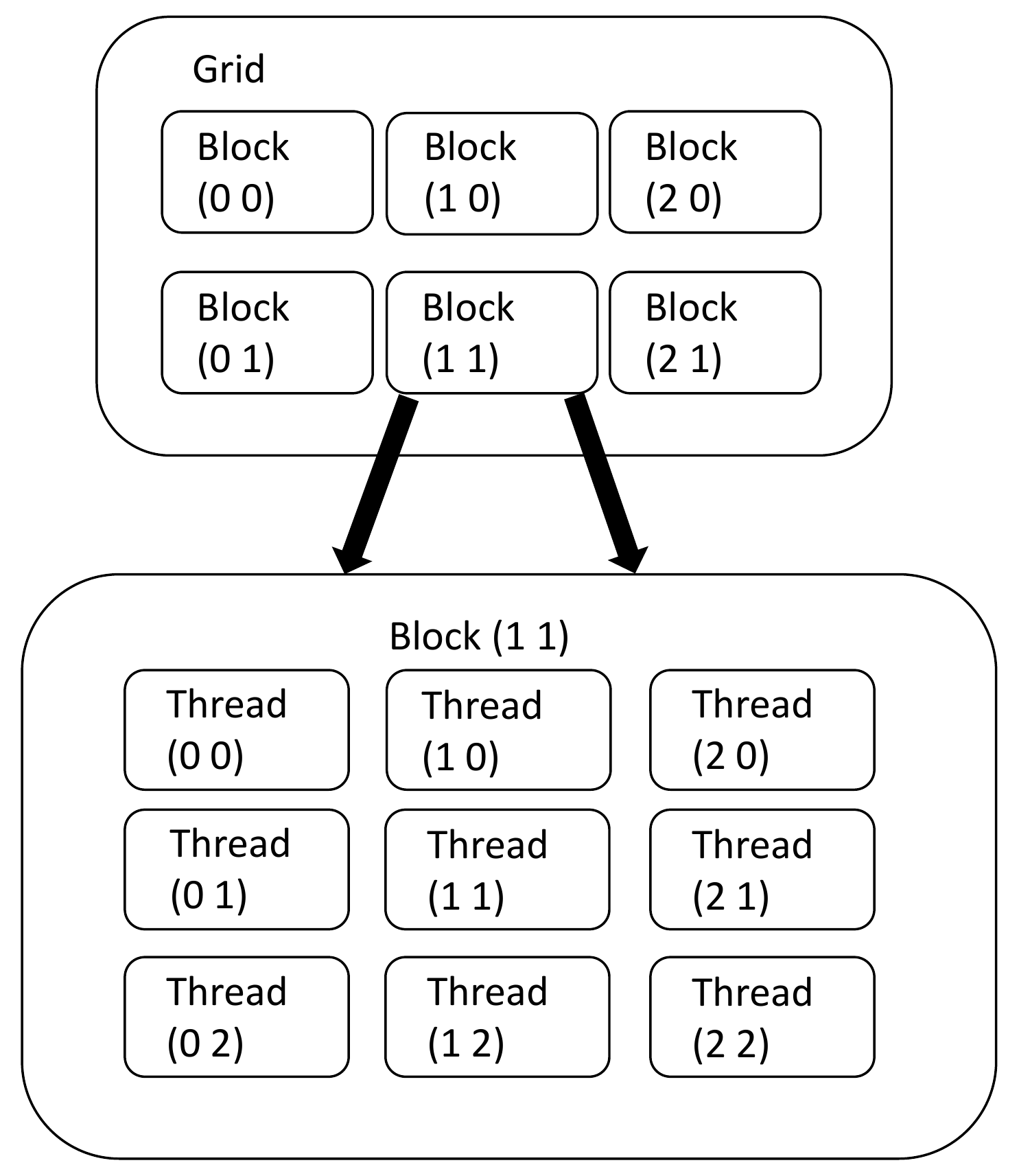}
 	\caption{CUDA Hierarchy\remove{The Hierarchy of CUDA Kernel Functions with Grids and Blocks}}
 	\label{fig_block}
 \end{figure}

Developers set dimensions of grids and blocks as inputs for executing their kernel functions. 
In particular, they divide computation into sub-computations and dispatch each sub-computation to different threads according to the grid and block dimensions. Eventually, the results of sub-computations can be merged as the final result of the overall computation through applying algorithms such as reduction. The hierarchy of the parallel computing mechanism of CUDA kernel functions is presented in Figure \ref{fig_block}.

To synchronize threads, CUDA applies barriers at which all the threads in one block must wait before any can proceed. In CUDA kernel functions, the barrier function is \codeIn{``\_\_syncthreads()''} which synchronizes threads from the same block. When a thread reaches a barrier, it is expected to proceed to next statement if and only if all threads from the same block have reached the same barrier. Otherwise, the program would be exposed to undefined behaviors.

\begin{figure}[t!]
	\centering
	\includegraphics[width=0.35\textwidth]{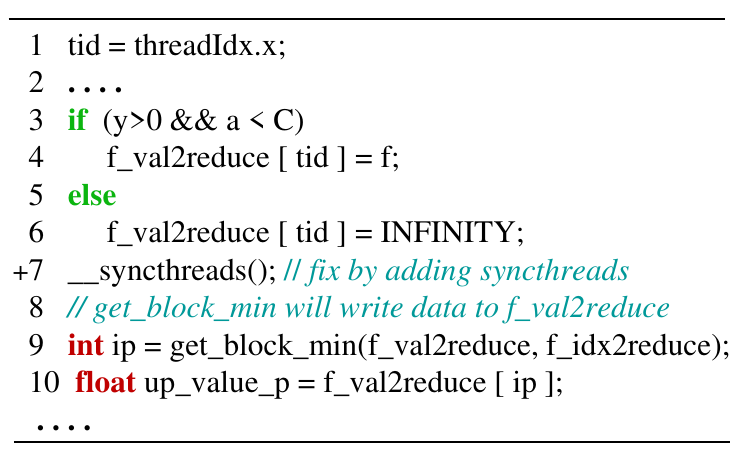}
	\caption{An Example of Data Race}
	\label{fig_section21}
\end{figure}

\textbf{CUDA Synchronization Bugs.} 
There are three major synchronization bugs in CUDA kernel functions: data race,  barrier divergence~\cite{CUDA2types}, and redundant barrier function based on our findings from our study in later sections. 
Specifically, data race indicates that for accessing global or shared memory, CUDA cannot guarantee the visit order of ``read\&write'' actions or ``write\&write'' actions from two or more threads.
For example, Figure~\ref{fig_section21} demonstrates the bug-fixing \commit{}
``\codeIn{febf515a82}'' in the file ``\codeIn{smo-kernel.cu}'' of the project ``\codeIn{thundersvm}''~\cite{thundersvm}, 
one of the highly-rated Github projects.
It can be observed from Figure~\ref{fig_section21} that the ``\codeIn{if}'' statement writes to the memory of ``\codeIn{f\_val2reduce}'', while inside the device, the function ``\codeIn{get\_block\_min}'' writes to the same memory. This ``write\&write'' bug is fixed by adding\remove{ the line of} ``\codeIn{\_\_syncthreads}'' which synchronizes actions among threads.

\begin{figure}
	\centering
	\includegraphics[width=0.35\textwidth]{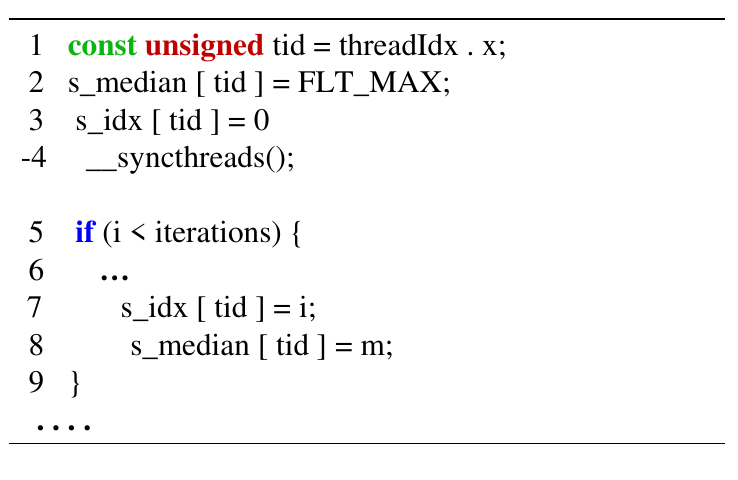}
	\caption{An Example of Redundant Barrier}
	\label{fig_section_new1}
\end{figure}

%

A barrier function is considered redundant when there is no data race after deleting it from source code. A redundant barrier function compromises the program performance in terms of time and memory usage. For instance, Figure~\ref{fig_section_new1} demonstrates the bug-fixing \commit{} ``\codeIn{31761d27f01}'' in the file ``\codeIn{kernel/homography.hpp}'' from the project ``\codeIn{arrayfire}''~\cite{arrayfire}. It can be observed that the block dimension is 1 since from Line 1, the value of ``\codeIn{tid}'' is assigned only by ``\codeIn{threadIdx.x}''. 
That indicates that the ``\codeIn{tid}''s are identical among different threads from the same block. As a result, ``\codeIn{s\_median[tid]}'' and ``\codeIn{s\_idx[tid]}'' can only be accessed by one thread, leading to a redundant barrier function in Line 4  because there is no race in ``\codeIn{s\_median}'' or ``\codeIn{s\_idx}'' after deleting it.

\begin{figure}
	\centering
	\includegraphics[width=0.35\textwidth]{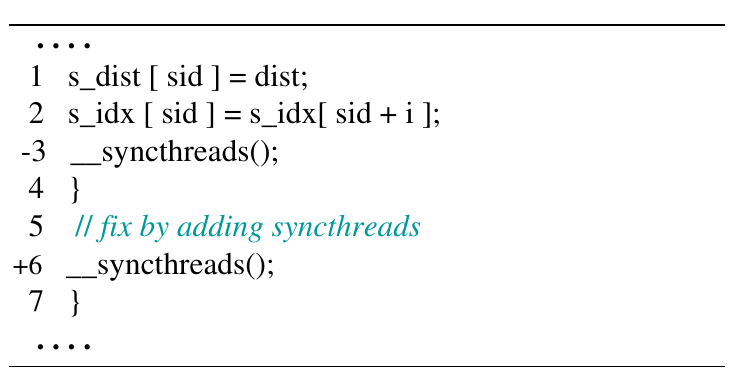}
	\caption{An Example of Barrier Divergence}
	\label{fig_section22}
\end{figure}

A barrier divergence takes place when some threads in a block complete their tasks and leave the barrier while the others have not reached the barrier yet. Figure~\ref{fig_section22} demonstrates the bug-fixing \commit{} ``\codeIn{0ed6cccc5ff}'' in the file ``\codeIn{nearest\_neighbour.hpp}'' from the project ``\codeIn{arrayfire}'' caused by barrier divergence.
It can be indicated from Figure~\ref{fig_section22} that developers make sure all the threads in the same block reach the same barrier in every execution of the kernel function by moving the statement of ``\codeIn{\_\_syncthreads()}'' outside the given branch. Otherwise they will have to handle undefined behaviors.

\section{Empirical Study}
\label{sec_study}
To investigate CUDA bug patterns, we first conduct a large-scale real-world bug dataset from open-source CUDA projects and then empirically analyzes the runtime stages and root causes of the collected CUDA program bugs. 
\subsection{Data collection and filtering}
To collect sufficient CUDA bugs for study, we first define policies to select  open-source CUDA projects. In this paper, we aim to select important and influential projects covering as many project types as possible. The collection is initialized by searching the keyword ``CUDA'' and results in more than 12,000 projects from \git{}. Next, we sort these projects in terms of the star number and commit number. 
In particular, the chosen projects fall into two groups: more active projects which are defined as the ones with over 1000 commits and less active projects which are defined as the ones with 500 to 1000 commits. 
In each group, we collect one library-type project and one application-type project with the most stars. However, none of these selected projects are marked with CUDA as their main language. Therefore, we collect one additional project with the main language marked as CUDA and most stars. As a result, we collect ``\codeIn{kaldi}''~\cite{kaldi} (more active applicatoin), ``\codeIn{arrayfire}''~\cite{arrayfire} (more active library), ``\codeIn{thundersvm}''~\cite{thundersvm} (less active application), ``\codeIn{mshadow}''~\cite{mshadow} (less active library), and ``\codeIn{cuda-convnet2}''~\cite{cudaconvnet2} (main laguage CUDA) as listed as in Table \ref{tab_study_project}.

After collecting the projects, the bugs are delivered based on the commit messages and ``\codeIn{git diff}'' results. The specific operations are listed as follows. 
We first filter the commits and only keep the commits with the messages that contain at least one keyword in the set \{``fix'', ``error'', ``mem''\}, following prior study on other types of bugs~\cite{Zhang:2018:EST:3213846.3213866}. In this way, the previous versions of the selected commits might have a higher chance to contain bugs. 

However, the commit messages only with these keywords might not be relevant with CUDA bugs. Therefore, next, among the filtered commit messages, we further filter them according to whether they have at least a keyword in the set \{``\codeIn{\_\_global\_\_}'', ``\codeIn{\_\_device\_\_}''\} or match at least one regular expression in the set \{``\codeIn{cuda\textbackslash w+\textbackslash s*[(]}'', ``\codeIn{[\^{}$<$]$<<<$[\^{}$<$]}''\} with its parent node's ``\codeIn{git diff}'' results. 
To illustrate, ``\codeIn{\_\_global\_\_}'' is the modifier of kernel functions and ``\codeIn{\_\_device\_\_}'' is the modifier of the device functions that can be called by kernel functions.\remove{ Therefore, the ``\codeIn{git diff}'' results of the dynamic execution of kernel functions are associated with either ``\codeIn{\_\_global\_\_}'' or ``\codeIn{\_\_device\_\_}''.} ``\codeIn{cuda\textbackslash w+\textbackslash s*[(]}'' is designed in accordance with the information that the resource is prepared/released in host side before/after executing kernel functions. For instance, ``\codeIn{cudaMalloc((void **) \&host, sizeof(int) *100)}'' allocates a global 400-byte memory for kernel functions before execution; ``\codeIn{cudaFree(\&host)}'' releases the allocated memory for kernel functions after execution. ``\codeIn{[\^{}$<$]$<<<$[\^{}$<$]}'' is designed in accordance with the scenario that sets up the environment for kernel functions, e.g., ``\codeIn{function\_name$<<<$grid\_size, block\_size$>>>$(arguments)}''. All these regular expressions together deliver the complete life cycle of executing kernel functions such that all the bugs of the whole life cycle can be covered. 

\begin{table}
\centering
\caption{\label{tab_study_project}Subject Statistics\remove{\lingming{why only the first project is capitalized? why not using macro?}}}
\begin{adjustbox}{width=0.8\columnwidth}
\begin{tabular}{l|lll}
\hline
\bf{Projects} & \bf{Star Number} & \bf{Commit Number} & \bf{LoC}\\
\hline\hline
\bf{kaldi}&5143&8419&364K\\
\bf{arrayfire}&2499&5171&381K\\
\bf{thundersvm}&818&790&343K\\
\bf{mshadow}&966&894&16K\\
\bf{cuda-convnet2}&620&40&27K\\
\hline
\end{tabular}
\end{adjustbox}
\end{table}

We further manually review all the remaining commits after two rounds of filtering to remove any potential false positives.\remove{are referred to as \textit{the bugs that are studied in this paper}, because they are much likely to indicate the changes caused by program fixes. }
Due to the tedious and time-consuming manual inspection, all the selected CUDA projects are analyzed within the most recent 
1000 commits\remove{\lingming{I think we should remove ``filtered''?}} or all of them if there are fewer than 1000 commits. 
As a result, we collected a total of 319 real-world CUDA bugs.  
Note that since CUDA programs are numeral-computation-oriented, they and their bug patterns appear to be converged as stated in the following sections. To the best of our knowledge, we conduct the most extensive study for CUDA program bugs to date. 

\subsection{Bug Taxonomy}
To understand the features of CUDA program bugs, we investigate\remove{ the bug scenarios, what bugs appear to be, and how exactly bugs are caused. Therefore, in this paper,} CUDA program bugs in the following dimensions: 

\subsubsection{Runtime stage}
Runtime stages refer to the life-cycle stages of running CUDA programs, 
including host resource preparation, kernel function execution, and host resource retrieve, as mentioned in Section~\ref{sec_bak}. In particular, ``kernel function execution'' tends to be more vulnerable to bugs than other runtime stages by involving 217 bugs out of 319 in total (217/319 = 68\%), while ``host resource retrieve'' takes up 11\% by involving 34 bugs and ``host resource preparation'' takes up 21\% by involving 68 bugs. In this paper, we focus on studying the runtime stage ``kernel function execution'' because programs in this stage are GPU-specific. 

\subsubsection{Bug root cause}

The root causes of CUDA program bugs can be grouped into five categories as follows. The detailed statistics for the root causes and their corresponding bug symptoms can be found in Table~\ref{tab_kernel_bug}.

a. \textit{\Improper{}}. 
This root cause refers to the bugs triggered when utilizing and managing memory and GPUs improperly. Such root cause widely spreads in the life cycle of CUDA programs and is associated with all the bug symptoms except \sometime{}. In particular, it includes the buggy scenarios such as incorrect device resource allocation, memory leak, early device call reset, and unauthorized memory access. From Table~\ref{tab_kernel_bug}, we can observe that \textit{\improper{}} takes up 14\% (31/217) among all the root causes. We can also notice that \improper{} is the major root cause for \crash{} (i.e., \remove{22/(5+3+1+22) = }73\%) because the memory issues incurred by \improper{} can result in possible fatal errors to crash programs.

b. \textit{\Non{}}. 
This root cause refers to the implementation which accomplishes the functional requirements with lossy performance. It is often associated with \always{} and \perform{} for various reasons, e.g., improper data type, outdated library functions, branch divergence in kernel functions. \textit{\Non{}} takes up 8\% (17/217) among all the root causes.

c. \textit{\Cross{}}. 
This root cause refers to the ones that occur in any platform or any programming language, such as range-checking errors, inappropriate exception handling, scope errors, and  other implementations that cannot accomplish given functional requirements. This root cause also widely spreads across all the life cycle of kernel functions. It can be observed that \cross{} is the major root cause of both \always{} (94\%) and all the bugs (63\%).

d. \textit{Improper synchronization}. 
This root cause is based on three CUDA-specific synchronization issues: data race, barrier divergence, and redundant barrier functions. Data race refers to when multiple threads ``read\&write'' or ``write\&write'' to the same memory address at the same time, the kernel functions may return different results in multiple executions even under the identical environmental setups. Barrier divergence leads to undefined behaviors while threads in the same block cannot reach the same barrier function. Redundant barrier function refers to that no data race exists by removing barrier functions. Figures \ref{fig_section21}, \ref{fig_section_new1}, and \ref{fig_section22} can be referred to for better illustration. Note that other issues such as improper-implementation-caused synchronization are also included here. Overall, \textit{Improper synchronization} takes up 12\% (27/217) among all the root causes. Due to the nature of such bugs, \perform{} and \sometime{}~\cite{luo2014empirical} (i.e., the tests with non-deterministic outcomes) are the main bug symptoms.

e. \textit{\Port{}}. 
This root cause refers to the issues that relate to certain platform specifics, such as operating systems or hardware platforms. For instance, on vs2013, bulding ``\codeIn{mshadow}''~\cite{mshadow} needs one additional step before calling ``\codeIn{\_\_half2float}''; otherwise the building would fail (\commit{} ``\codeIn{51a8a7e3e5}'' of the project ``\codeIn{mshadow}'').  \Port{} is the most rare root cause for all the CUDA bugs (3\%).

\begin{table}
\centering
\caption{\label{tab_kernel_bug}Root Causes and Bug Symptoms for CUDA Kernel}
\begin{adjustbox}{width=\columnwidth}
\begin{tabular}{|l||llll|l|}
\hline
\bf{Root Causes} & \bf{\Crash{}} & \bf{\Always{}} & \bf{\Perform{}} & \bf{\Sometime{}} &\bf{Sum} \\  
\hline 
\hline
\bf{\Improper{}} & 22 & 2 & 7 & 0 & 31 \\ 
\bf{\Non{}} & 0 & 7 & 10 & 0 & 17 \\ 
\bf{\Cross{}}  & 3 & 132 & 1 & 0 & 136 \\
\bf{\textit{Improper synchronization}}  &0  & 0 & 10 & 17 & 27 \\
\bf{\Port{}}  &5  & 0 & 0 & 1 & 6  \\
\hline
\bf{Sum} &30 &141&28&18&217
\\
\hline
\end{tabular}

\end{adjustbox}
\end{table}

\subsection{Bug Impact and Detection Effort}
Our study derives the following findings on bug impact and detection effort.

\subsubsection{The majority of the kernel function bugs are not CUDA-specific or single-instruction-multiple-data (SIMD) only and can take place in other platforms and be detected by traditional approaches}

 As mentioned before, \cross{} is the major bug type among all. In particular, for the most of these bugs caused by \cross{} that do not exclusively arise only in CUDA, it is routine to design test cases similar for generic software programs. To conclude, it can be implied that to detect these bugs, we can use traditional approaches as for other program types without designing new approaches~\cite{Jones:2005:EET:1101908.1101949, 6227211}.


\subsubsection{The majority of the memory-related bugs can be solved by traditional CPU-based approaches}

In this paper, the memory-related bugs mainly refer to \crash{} bugs and the others caused by \improper{}. Among the total 36 bugs regarding global and shared memory, memory leak (13) and unauthorized memory access (20) are the major ones.

Most of the CUDA programs rely on the injected grid or block dimensions to determine the thread-wise memory-access range. If there are computing errors or improper memory-access ranges, it tends to cross borders and leads to incorrect programming outputs or even program crashes. To detect such bugs, when kernel functions are launched, it is applicable to determine whether a statement has cross-border access according to whether the thread-wise memory-access range of that statement are intertwined or violates the preset legitimate global memory-access range, e.g., \commit{} ``\codeIn{97cca6c0ff6}'' from the project ``\codeIn{kaldi}''\remove{\lingming{why kaldi is not capitalized here but capitalized in later tables?}} and \commit{} ``\codeIn{ba19743bb6}'' from the project ``\codeIn{arrayfire}''. Some CPU-based approaches, such as~\cite{7194597}, can be studied for resolving such problems.

\subsubsection{Detecting synchronization bugs is important and time consuming}

We believe synchronization bugs have significant impacts on CUDA programs according to the following four reasons: 
(1) the occurrence of the synchronization bugs is non-negligible (27/217 = 12.4\%); (2) such bugs are hard to be reproduced and could increase the difficulties of testing and debugging (shown in Table~\ref{tab_kernel_bug}, such bugs usually incur performance issues and flaky tests, both of which are hard to diagnose); (3) the synchronization bugs are tightly connected with CUDA specifics that are hard to be detected by the existing CPU-based approaches, while there exist traditional approaches can be adopted for detecting the majority of the other bugs, e.g., \cross{}; (4) some synchronization bugs, e.g., data race, can easily taint the computation results or even shelter other bug types to render them more challenging to be captured.

\begin{figure}
\includegraphics[height=3cm]{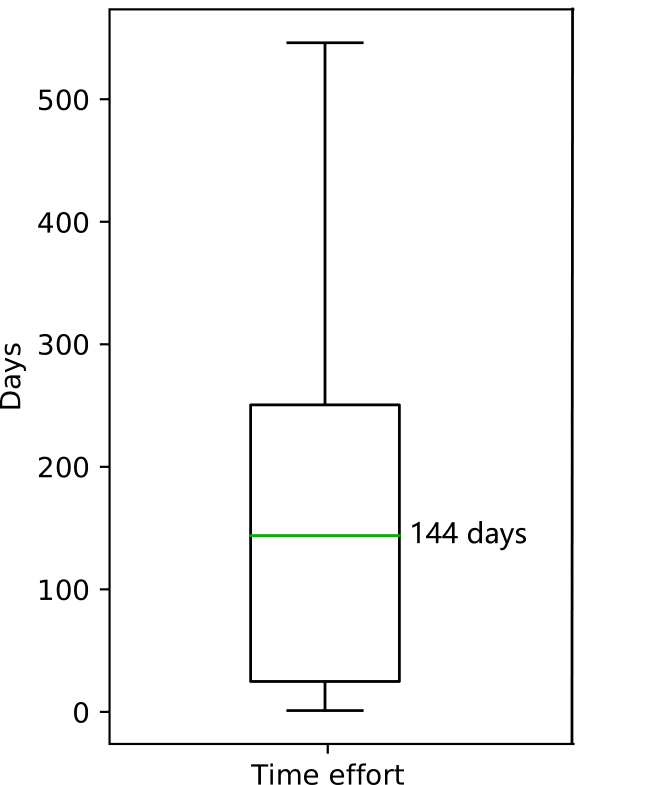} \centering
\caption{Synchronization bug impacts}
\label{fig_effort}
\end{figure}
In this paper, the impacts of synchronization bugs are measured by \emph{time effort} which is defined as the time window between the commit where the buggy code was firstly introduced and the commit where the bug was fixed\remove{that the commit message of the corresponding bug being fixed subtracting the time that the commit message of the relevant changes of the bug-induced code being made}. 
From Figure~\ref{fig_effort}, it can be observed that a number of ``\sometime{}'' synchronization bugs have been existing in the program for a long time (i.e., median 144 days, averagely 180 days, 11 bugs out of 17 over 100 days). 
Therefore, it can be concluded that the synchronization bugs are hard to be detect and fix. 

\vspace{0.2cm}

\fbox{%
    \parbox{0.9\columnwidth}{%
       \textit{Among all the root causes regarding CUDA kernel functions, \sync{} is challenging and time-consuming to be detected and requires new detection approaches.}
    }%
}

\subsection{Bug Detection Motivation and Possibility}
It can be inferred from the previous study that synchronization bugs have a significant impact on executing CUDA kernel functions. Recently, some compiler-based approaches have been proposed to detect CUDA synchronization bugs, such as  \curd{}~\cite{Peng:2018:CDC:3296979.3192368},  \barracuda{}~\cite{barracuda}. Typically, these approaches link the detectors to the applications in the compiling stage and check the bugs in runtime process. However, they are limited by not being ``fully automatic''---users have to provide error-inducing inputs manually. On the other hand, given inferior inputs, the synchronization bugs might not be triggered and detected. Moreover, it can be expensive since such runtime detection demands compiling process and GPU computing environment. Other automatic synchronization bug detection approaches, e.g., \gklee{}~\cite{gklee}, apply static/dynamic analysis and could lead to poor runtime performance on real-world projects. In addition, most of them~\cite{Li2014LDetectorAL}~\cite{racecheckor} are designed only for certain bug types, e.g., data race, while they are limited in detecting other bug types (e.g., barrier divergence and redundant barrier function). Therefore, it is essential to develop a new approach to efficiently detect a wide range of synchronization bugs.

It can be observed from our study that traditional test cases are able to capture various bug types except CUDA synchronization bugs. On the other hand, it can be inferred that with ideal test cases, synchronization bugs are expected to be captured easily. Such test cases should differ from traditional test cases which only deliver runtime program output. Instead, they should deliver the information that can help capture the synchronization occurrence, e.g., the deterministic thread-wise memory visit order. On this purpose, considering that it is hard to capture such information by directly running CUDA programs, a possible idea is to design a virtual machine that can offload runtime GPU programming to offline CPU programming for better observing runtime information in kernel functions and set a mechanism to collect and analyze them. Accordingly, error-inducing grids and blocks based on their contexts can be approached in a fully automated manner to detect synchronization bugs.

This idea is applicable due to the following reasons: (1) the sophisticated parallel computing model of CUDA allows collecting various runtime information without reducing significant runtime performance, and (2) the entire process is essentially simulating runtime CUDA program with collecting runtime information which is expected to be as closely efficient as simply running the original programs. 

\begin{figure}
	\centering
	\includegraphics[width=0.3\textwidth]{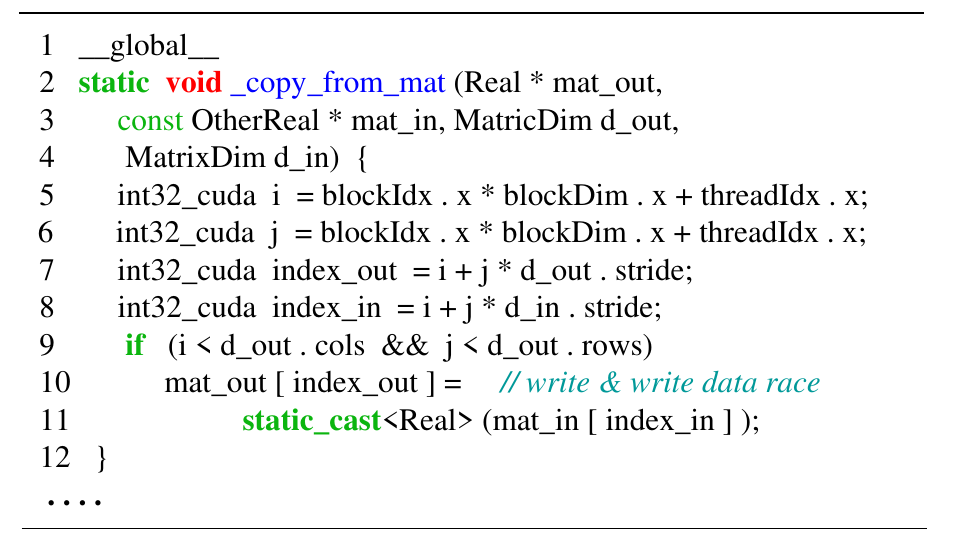}
	\caption{An Example to Illustrate the Possibility to Automatically Detect Bugs}
	\label{figexample}
\end{figure}

Figure \ref{figexample} is an example from the commit of latest version of the project ``\codeIn{kaldi}''. 
For this bug, normally the developers have little clue about whether this piece of code has synchronization bugs because regular test cases may pass during many runs.
Our intuition in designing specific test cases for synchronization bugs might detect synchronization bugs in this case by
setting up initial environment for ``\codeIn{\_copy\_from\_mat}''. If they set the grid dimension to be 1, the block dimension to be (3, 2), ``\codeIn{d\_in.stride}'' to be 1, ``\codeIn{d\_out.stride}'' to be 1, ``\codeIn{d\_out.rows}'' to be 5, and ``\codeIn{d\_out.cols}'' to be 5, then thread (0 1 0) and thread (1 0 0) will report ``write \& write'' data race at Line 10, indicating that when ``\codeIn{d\_out.stride}'' is smaller than ``\codeIn{d\_out.cols}'', kernel functions should instantly raise an exception that ``\codeIn{stride}'' should be larger than ``\codeIn{cols}'' instead of executing all the code for a long time. Eventually, the problem of automatically detecting such synchronization bugs can be transformed to the problem of automatically generating the error-inducing grid and block dimensions.\\

\fbox{%
    \parbox{0.9\columnwidth}{%
       \textit{It is possible to design a general automated framework to detect CUDA synchronization bugs as long as tests (e.g., error-inducing grid and block dimensions) can be automatically approached to collect and analyze the corresponding memory-access information.}
    }%
}

\section{Framework of Simulee}
\label{sec_gunit}

\begin{figure}[h]
\includegraphics[width=8cm,height=5.3cm]{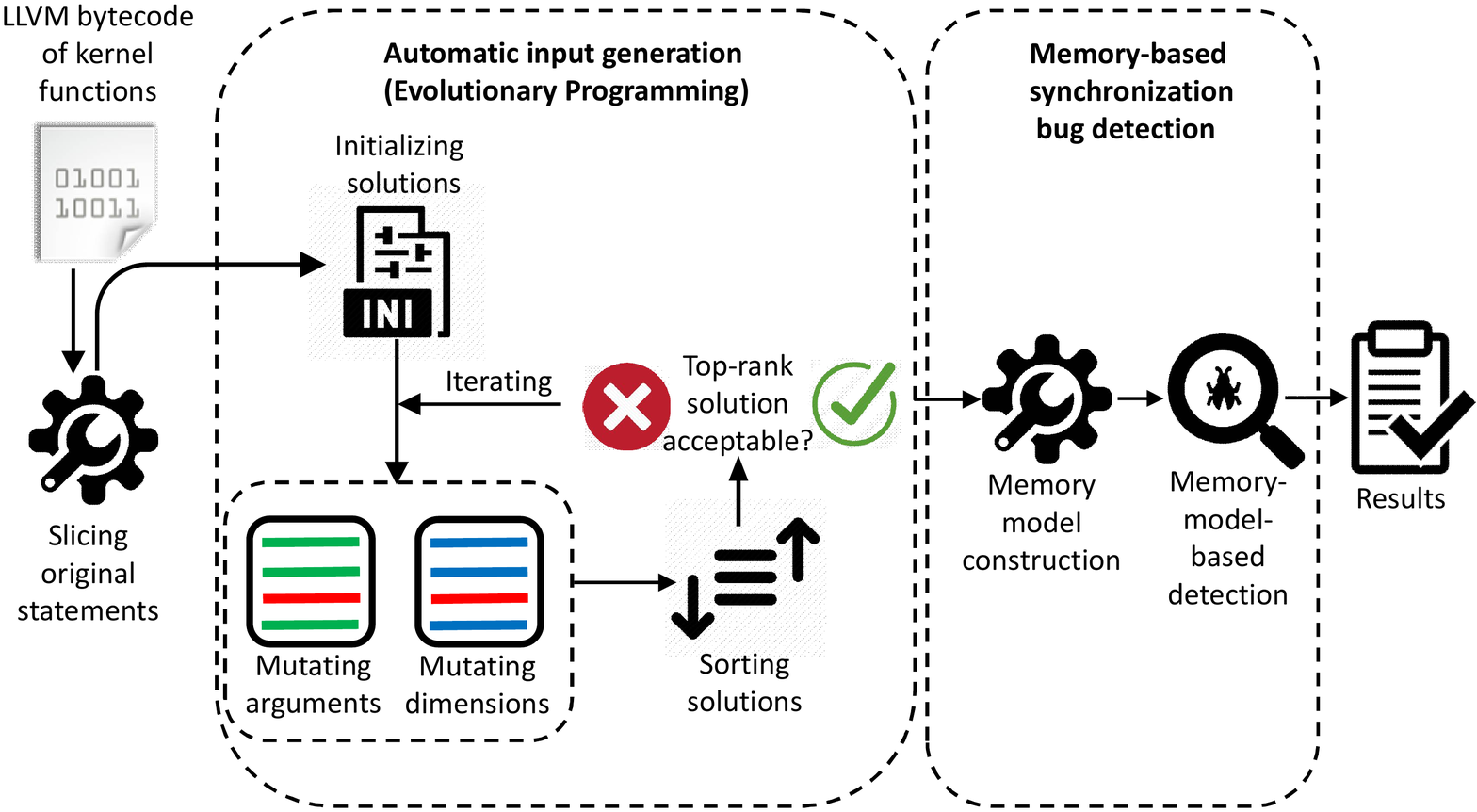}
\centering
\caption{Framework of \simulee{}}
\label{fig_workflow}
\end{figure}

In this section, we introduce \simulee{}, an automatic tool to detect real-world CUDA synchronization bugs. Typically, \simulee{} takes llvm bytecode 
translated from CUDA kernel function programs, automatically generates the associated error-inducing inputs, and yields \model{} to detect synchronization bugs. Specifically, \simulee{} is composed of two parts---``Automatic Input Generation'' and ``\model{}-based Bug Detection''. ``Automatic Input Generation" is initialized by inputting the llvm bytecode of CUDA kernel function programs. Next, it slices the memory-access statements (e.g., read and write statements) and inputs them for Evolutionary Programming~\cite{evolutionaryprogramming}. 
Subsequently, Evolutionary Programming helps generate error-inducing environmental setups by iteratively mutating and sorting dimensions and arguments and passes the acceptable ones to ``\model{}-based Bug Detection''. At last, ``\model{}-based Bug Detection'' simulates runtime environment by constructing \model{} and using it to detect whether there are synchronization bugs. 
The details can be referred to in Figure \ref{fig_workflow}.

\subsection{Automatic Input Generation}
Generating error-inducing inputs is essentially equivalent to generating the inputs that can lead to the memory-access conflicts among threads to improve the possibility of CUDA synchronization bug occurrences. However, how to automatically generate such error-inducing inputs remains challenging. Some intuitive solutions, such as random generation or coverage-oriented generation might be limited in effectiveness and efficiency, since they are not specially designed for triggering memory-access conflicts. In this section, we introduce how \simulee{} automatically generates error-inducing inputs for detecting CUDA synchronization bugs in an effective and efficient manner.



\subsubsection{Intuition}
An effective and efficient automatic approach to generate error-inducing inputs for triggering CUDA synchronization bugs indicates to generate as many memory-access conflicts as possible within a short time limit. 
Given the \textit{i}th memory address and the kernel function inputs, i.e., grid and block dimensions and arguments, \heuristic{} is defined as the number of threads that access the \textit{i}th memory address while \textit{g(i)} is a function that returns 1 when the \textit{i}th memory address is accessed by any thread and returns 0 otherwise. [\textit{start}, \textit{end}] denotes the memory-access range.  An intuitive target function \targetFunction{} can be presented in Equation \ref{intuitionformula} which denotes the ratio of the total number of the accessed memory addresses to the total number of the  memory-access threads: 
\begin{equation}
\small
\label{intuitionformula}
F(dimensions, arguments)=\frac{\displaystyle\sum_{i=start}^{end} g(i)}{\displaystyle\sum_{i=start}^{end} f(i)}
\end{equation}
It can be derived that the max value of \targetFunction{} is 1 which denotes that there is no memory-access conflict between any thread pair. On the other hand, the smaller \targetFunction{} is, the higher chance the memory-access conflict takes place. 
Therefore, \targetFunction{} can be used for optimization to obtain error-inducing inputs that trigger CUDA synchronization bugs. 
Note that since \targetFunction{} is discrete, we choose Evolutionary Programming~\cite{evolutionaryprogramming} as our optimization approach. 

\subsubsection{Framework}
The framework of ``Automatic Input Generation'' is presented in Algorithm \ref{alg_evolution}. 
First, \simulee{} randomly initializes \textit{arguments} and \textit{dimensions} to create and sort individual solutions for evolving (Lines 3 to 7). In each generation, each solution is mutated to generate two children, which are added to the whole population set (Lines 8 to 14). Next, the population winners survive for the subsequent iterations (Lines 15 to 16). 
The iterations can be terminated once it finds an acceptable solution. Otherwise, after completing the iterations, it returns the optimal solution by then.

\textbf{\textit{Initial Solutions. }}The initial dimensions and arguments are randomly generated and passed to fitness functions as initial solutions for future evolution. Note that the dimensions can be extracted from kernel functions. For instance, if a kernel function has  ``threadIdx.x'' and ``threadIdx.y'', it means the block is two-dimensional.

\textbf{\textit{Fitness Function. }}Equation \ref{intuitionformula} is chosen as the primary fitness function for Evolutionary Programming. Specifically, the output of \targetFunction{} is the fitness score for a solution of dimensions and arguments in Evolutionary Programming. 
However, it is difficult to derive an optimal solution of dimensions and arguments by only optimizing \targetFunction{}. In particular, since \targetFunction{} is non-differentiable when the gradient does not exist, it is hard to find an optimal solution given the set of inferior solutions, e.g., all the solutions of \targetFunction{} are ``1''s. 
To tackle such inferior solutions, we design a secondary fitness function such that they are sorted according to their possibility to be optimal: $R(start, end) = end - start$.\remove{, as presented in Equation \ref{secondary}.} In particular, it indicates that a smaller memory-access range leads to a higher possibility of memory-access conflict. As a result, we define fitness score of the primary fitness function as \first{}, and the fitness score of the secondary fitness function as \second{}.
During the population evaluation, the \first{} is sorted first; if and only if the top-ranked \first{} is 1, the \second{} is sorted to decide which solution is more likely to converge to the minimum of \targetFunction{}.

\begin{algorithm}[!t]
	\caption{Framework for Automatic Input Generation}
	\label{alg_evolution}
	\begin{flushleft}
		\hspace*{\algorithmicindent} \textbf{Input }:  population, generation
	\end{flushleft}
	\begin{flushleft}
		\hspace*{\algorithmicindent} \textbf{Output}: acceptable arguments and dimensions
	\end{flushleft}
	\small{
		\begin{algorithmic}[1]
			\Function{EVOLUTION\_ALGORITHM}{}
			\State population\_lst $\leftarrow$ list()
			\For{i in population}
			\State single\_solution $\leftarrow$ InitialSolution()
			\State single\_score $\leftarrow$ fitness(single\_solution)
			\State population\_lst.append([single\_solution, single\_score])
			\EndFor
			\State sort\_by\_score(population\_lst)
			\For{i in generation}
			\State child\_lst $\leftarrow$ list()
			\For{solution in population\_lst}
			\State children\_solutions $\leftarrow$ mutation(solution)
			\State new\_scores $\leftarrow$ fitness(children\_solutions)
			\State child\_lst.append([children\_solutions, new\_scores])
			\EndFor
			\State population\_lst.merge(child\_lst)
			\State sort\_by\_score(population\_lst)
			\State population\_lst $\leftarrow$ population\_lst[:population]
			\If{population\_lst[0] acceptable}
			\State \textbf{return} population\_lst
			\EndIf
			\EndFor
			\State \textbf{return} population\_lst
			\EndFunction
		\end{algorithmic}
	}
\end{algorithm}

\textbf{\textit{Mutation. }} In \simulee{}, solutions are generated by mutation, where each solution generates two children in one generation. 
Specifically, \textit{arguments} and \textit{dimensions} are independent from each other during mutation with respective mutation strategies. 
The mutate strategy for \textit{dimensions} is trivial: first, \simulee{} randomly generates an integer vector ranging from -1 to 1 according to the dimension size; next, the child's dimension is mutated by summing the parent's dimension and the generated integer vector. 

The details of the mutation strategy for \textit{arguments} is presented in Algorithm \ref{alg_arguments}. Since the memory-access-relevant arguments are numbers, \simulee{} considers them as float numbers and converts them back to the actual types when executing \heuristic{}. Accordingly, each generation generates two children: one adds a random number generated by standard Normal Distribution~\cite{normalwiki} $(N(x) = \frac{1}{{ \sqrt {2\pi } }}e^{{{ - x^2 } \mathord{\left/ {\vphantom {{ - \left( {x - \mu } \right)^2 } {2}}} \right. \kern-\nulldelimiterspace} {2}}})$ to the arguments inherited from the parent solution, and the other adds a random number generated by standard Cauchy Distribution~\cite{cauchywiki} $(C(x) = \frac{1}{\pi (1 + x^2)})$ to the arguments inherited from the parent solution. 
We define the search step length of the arguments as the absolute value of the number generated from the two aforementioned distributions, with expected values shown in Equations \ref{normal} and \ref{cauchy}. 
\begin{equation}\small
\label{normal}
E_{normal}(x)=\int_{0}^{\infty} x\frac{1}{{ \sqrt {2\pi } }}e^{{{ - x^2 } \mathord{\left/ {\vphantom {{ - \left( {x - \mu } \right)^2 } {2}}} \right. \kern-\nulldelimiterspace} {2}}}dx = 0.399
\end{equation}
\begin{equation}\small
\label{cauchy}
E_{cauchy}(x)=\int_{0}^{\infty} x\frac{1}{\pi (1 + x^2)} dx = +\infty
\end{equation}
We next explain why we apply the above two distributions.
It can be observed from Equations \ref{normal} and \ref{cauchy} that, the step length generated from standard normal distribution is expected to be small. That indicates that if there is an optimal solution nearby, the generated child is likely to approach it. On the contrary, the step length generated from standard cauchy distribution is expected to be large. That indicates that if there is an inferior solution nearby, the  generated child is likely to escape from it.

\textbf{\textit{Acceptable Function.}} The acceptable function is used to terminate the whole process given an acceptable solution. In our work, the acceptable solution is defined as that \first{} is smaller than 0.3.

To conclude, by applying Evolutionary Programming, \simulee{} is expected to deliver error-inducing grid and block dimensions and arguments that lead to memory-access conflicts and trigger CUDA synchronization bugs.

\begin{algorithm}[t!]
	\caption{Mutating Arguments}
	\label{alg_arguments}
	\begin{flushleft}
		\hspace*{\algorithmicindent} \textbf{Input }:  parent
	\end{flushleft}
	\begin{flushleft}
		\hspace*{\algorithmicindent} \textbf{Output}: normal\_solution, cauchy\_solution
	\end{flushleft}
	\small{
		\begin{algorithmic}[1]
			\Function{ARGUMENT\_MUTATION}{}
			\State normal\_solution $\leftarrow$ copy(parent)
			\State cauchy\_solution $\leftarrow$ copy(parent)
			\For{ argument in parent}
			\State normal\_solution[argument] $\leftarrow$ parent[argument] + normal()
			\State cauchy\_solution[argument] $\leftarrow$ parent[argument] + cauchy()
			\EndFor
			\State \textbf{return} normal\_solution, cauchy\_solution
			\EndFunction
		\end{algorithmic}
	}
\end{algorithm}

\subsection{Memory-based Synchronization Bug Detection}
With the auto-generated error-inducing inputs, the synchronization bug detection of \simulee{} is established on building a \model{} that depicts thread-wise memory-access instances. Based on the \model{}, \simulee{} develops a set of criteria to detect synchronization bugs including data race, redundant barrier functions, and barrier divergence. 

\subsubsection{Memory Model}
The \model{} accessed by the kernel functions is defined to be composed of a set of \unit{}s where each \unit{} corresponds to a memory address and is composed of a set of \tuple{}s. A \tuple{} is defined as a three-dimensional vector space $<$\vo{}, \ti{}, \ac{}$>$, where \vo{} represents the visit order to the associated memory address from different threads, \ti{} represents the indices of such threads, and \ac{} refers to the read or write action from those threads.

\begin{figure}
	\includegraphics[width=0.3\textwidth]{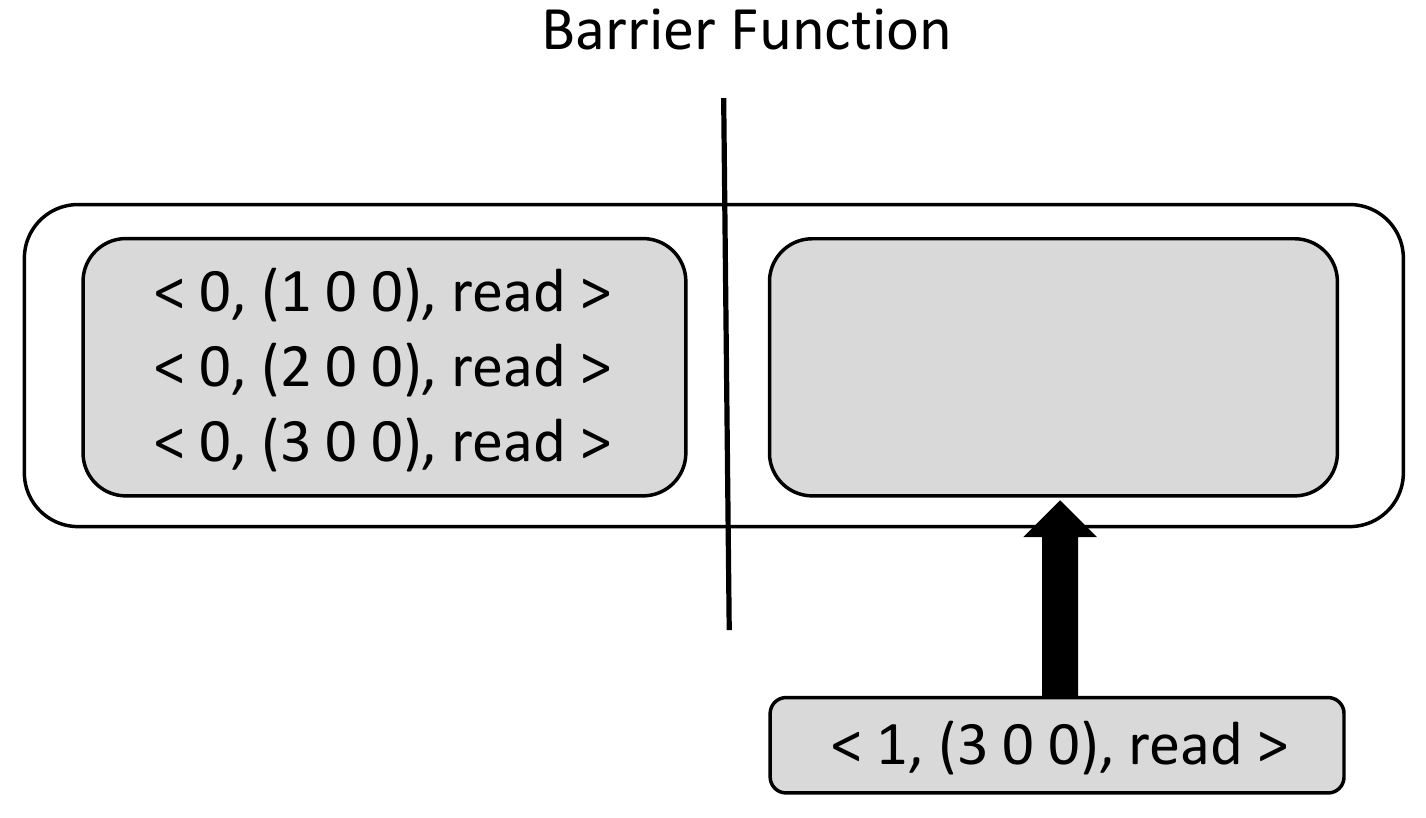}
	\centering
	\caption{\unit{} Example}
	\label{fig_torder2}
\end{figure}

An example of \unit{} is demonstrated in Figure \ref{fig_torder2} with four \tuple{}s $<$0, (1 0 0), read$>$, $<$0, (2 0 0), read$>$,
$<$0, (3 0 0), read$>$, and $<$1, (3 0 0), read$>$ where threads (1,0,0), (2,0,0), and (3,0,0) read the same memory address in the same \vo{} since none of them have reached any barrier function before they read. Assume all the threads reach a barrier function later and thread (3 0 0) reads, the \vo{} is then incremented from 0 to 1 for thread (3 0 0) and the other threads afterwards. 

\subsubsection{Memory Model Construction}
\label{construction}
Since \model{} is only associated with barrier functions and memory-access statements, it is applicable to detect synchronization bugs by obtaining such statements and then extracting/analyzing the memory-access information instead of executing the complete CUDA programs on GPUs, i.e., \textbf{simulating} the execution of CUDA kernel function programs. This simulation process is initiated by inputting the auto-generated block and grid dimensions and arguments passed to the kernel functions. Next, it constructs the \unit{} for each memory address.

The overall \model{} construction is demonstrated in Algorithm \ref{alg_cedf}. In particular, the algorithm is launched to initialize the block and grid dimensions as well as the global and shared memory for each thread (Lines 2 to 5). Next, for each block, the shared memory (Line 7) and the thread-wise \vo{} for each global and shared memory address (Lines 8 to 9) are initialized. If there are still some unterminated threads, for all of them, their corresponding \unit{}s are derived based on the collected parameters, e.g., \codeIn{global\_mem} and \codeIn{visit\_order\_global} (Lines 10 to 14). The construction of the thread-wise \unit{}s for shared memory and global memory are completed if there is no running thread left (Lines 15 to 16). 

Algorithm \ref{alg_threads} illustrates the details of \model{} construction for a single thread. Specifically, given a running thread and the parameters passed by Algorithm~\ref{alg_cedf} (Lines 2 to 4), Algorithm \ref{alg_threads} is initialized by detecting whether the current statement is the end of file. If so, the thread would be terminated. If there is any thread halting afterwards, we can confirm there is a ``barrier divergence'' bug because that indicates at least a thread has not reached the barrier function where the other threads of the same block all have completed their tasks and left (Lines 5 to 9).

If the current statement calls barrier function and all the other threads have reached the same barrier function, the \vo{} for both global and shared memory would be incremented if they have been visited before (Lines 10 to 13), since it indicates that all the threads in one block have visited the current memory address and the subsequent visits would demand a new barrier function. On the other hand, if the current statement does not call barrier function, the corresponding \vo{} and the \ac{} of the associated thread is recorded to construct the \model{} (Lines 14 to 21).  


\begin{algorithm}[!t]
  \caption{\model{} construction}
  \label{alg_cedf}
  \begin{flushleft} 
  \hspace*{\algorithmicindent} \textbf{Input }:  grid\_dim, block\_dim, arguments
  \end{flushleft}
  \begin{flushleft} 
  \hspace*{\algorithmicindent} \textbf{Output}: \model{}
  \end{flushleft}
  \small{
	\begin{algorithmic}[1]
	

    \Function{CONSTRUCT\_MEMORY\_MODEL}{}
        \State BLOCKS, $\leftarrow$
        generate\_from\_dimension(grid\_dim)
        \State THREADS $\leftarrow$ generate\_from\_dimension(block\_dim)
        \State global\_mem $\leftarrow$ [MemoryUnit() for i in range(global\_size)]
        \State shared\_mem\_lst $\leftarrow$ list()
        \For{blk in BLOCKS}
        \State shared\_mem $\leftarrow$ [MemoryUnit() for i in range(shared\_size)]
        \State visit\_order\_global $\leftarrow$ [0 for i in range(global\_size)]
        \State visit\_order\_shared $\leftarrow$ [0 for i in range(shared\_size)]
        \While{has\_unterminated\_thread()}
        \For{t in THREADS}
        \State env $\leftarrow$ Environment(arguments)

        \State PROCESS\_THREAD(t, global\_mem, shared\_mem,
        \State visit\_order\_global, visit\_order\_shared, env)
        \EndFor
        \EndWhile
        \State shared\_mem\_lst.append(shared\_mem)
        \EndFor
        \State \textbf{return} global\_mem, shared\_mem\_lst
    \EndFunction
	\end{algorithmic}
    }
\end{algorithm}

\subsubsection{Memory-Model-based Detection Mechanism}
The design of \model{} can be used in \simulee{} to detect CUDA synchronization bugs, i.e., data race, redundant barrier function, and barrier divergence. 

\textbf{\textit{Data Race.}} In general parallel computing programs, a possible data race takes place when multiple threads access the identical memory address in the same visit order and at least one of them writes. Specifically in CUDA kernel functions, besides the generic circumstances, a data race also takes place when (1) the threads are from different thread warps, or (2) the threads from the same thread warp underwent branch divergence, or (3) the threads from the same thread warp without undergoing branch divergence write to the same memory address by the same statement.  
By combining the data race detection criteria above and the design of \model{}, \simulee{} can detect data race in CUDA kernel functions as described in Theorem \ref{datarace}.
\begin{theorem}
Given two \tuple{}s $\psi_{i}$ and $\psi_{j}$ from the identical \unit{}, a data race between them takes place if the conditions below are met:
\begin{itemize}
    \item $\psi_{i}$[\vo{}] = $\psi_{j}$[\vo{}]
    \item $\psi_{i}$[\ti{}] != $\psi_{j}$[\ti{}]
    \item $\psi_{i}$[\ac{}] = `write' or $\psi_{j}$[\ac{}] = `write'
\end{itemize}
when the threads of $\psi_{i}$ and $\psi_{j}$ are (1) from different thread warps or (2) executing the ``write'' action on the same statements in the same thread warp or (3) underwent branch divergence before the current ``write'' action. 

\label{datarace}
\end{theorem}

\textbf{\textit{Redundant Barrier Function. }}
A redundant barrier function indicates that no data race can be detected by removing that barrier function. 
In CUDA kernel functions, the \vo{} is incremented for one \tuple{} when at least one thread reaches a barrier function. In other words, two \tuple{}s with adjacent \vo{} in one \unit{} indicates the presence of a barrier function, shown in Figure~\ref{fig_torder2}. Therefore, to detect whether a barrier function is redundant or not, it is essential to collect all the associated \tuple{}s and analyze whether they together would lead to data race. The barrier function is defined to be redundant if no data race can be detected among such \tuple{}s. 

The details of how to detect data race and redundant barrier function based on \model{} are presented in Algorithm \ref{alg_detect}. For each \unit{}, to detect data race, \simulee{} first groups the \tuple{}s with the same \vo{}. For all the \tuple{}s in one group, \simulee{} checks whether any \tuple{} has data race with others according to Theorem \ref{datarace} (Lines 4 to 16). To detect redundant barrier function of one \unit{}, \simulee{} extracts its \vo{} and groups all the \tuple{}s with adjacent \vo{} to find out whether any data race can take place (Lines 18 to 22). If there is no data race, \simulee{} identifies the associated barrier function and increments its recorder by 1 (Lines 23 to 24). At last, it checks whether the total recorder number matches the total number of the changing \vo{} caused by that barrier function which can be obtained after constructing the \model{}. This barrier function is redundant if the two numbers are equivalent (Lines 25 to 28).  

\textbf{\textit{Barrier Divergence. }}
As mentioned in Section~\ref{construction}, barrier divergence can be detected during constructing \model{} when there is any halting thread after the current execution is terminated, because it indicates that there is at least one thread which has not reached the barrier function while the others have already left. 

To conclude, \simulee{} first applies Evolutionary Programming to generate error-inducing grid and block dimensions and arguments. Next, \simulee{} inputs such dimensions and arguments to construct \model{} that delivers thread-wise memory-access information. Eventually, such information, along with the CUDA synchronization bug detection mechanism, are used to detect whether there exists any CUDA synchronization bug. 

\begin{algorithm}[t!]
  \caption{Thread Processor}
  \label{alg_threads}
  \begin{flushleft}
  \hspace*{\algorithmicindent} \textbf{Input }:  thread, global\_mem, shared\_mem, visit\_order\_global,
  \hspace*{\algorithmicindent} 
  \hspace*{\algorithmicindent}
  \hspace*{\algorithmicindent}visit\_order\_shared, env 
  \end{flushleft}
  \begin{flushleft}
  \hspace*{\algorithmicindent} \textbf{Output}:None or BARRIER\_DIVERGENCE
  \end{flushleft}
  \small{
	\begin{algorithmic}[1]
    \Function{PROCESS\_THREAD}{}
        \If{should\_halt() or is\_finished()}
        \State \textbf{return}
        \EndIf
        \State cur\_stmt $\leftarrow$ env.get\_next\_IP()
        \If{cur\_stmt.is\_EOF()}
        \State thread.finish()
        \If{has\_halt\_threads()}
        \State \textbf{return} BARRIER\_DIVERGENCE
        \EndIf
        \State \textbf{return}
        \EndIf
        \If{cur\_stmt.is\_syncthreads()}
        \If{all threads reach same barrier}
        \State update\_current\_visit\_order(visit\_order\_shared)
        \State update\_current\_visit\_order(visit\_order\_global)
        \EndIf
        \Else
        \State is\_global, mem\_index $\leftarrow$ simulate\_execute(cur\_stmt, env)
        
        \If{is\_global}
        \State index $\leftarrow$ visit\_order\_global[mem\_index]
        \State update\_memory\_model(global\_mem, mem\_index, index)
        \Else
        \State index $\leftarrow$ visit\_order\_shared[mem\_index]
        \State update\_memory\_model(shared\_mem, mem\_index, index)
        \EndIf
        \EndIf
        \State \textbf{return}
    \EndFunction
	\end{algorithmic}
    }
\end{algorithm}

\remove{The mechanism to detect barrier divergence would be discussed in Section \ref{construction}. }

\begin{algorithm}[!t]
  \caption{\model{}-based Detection}
  \label{alg_detect}
  \begin{flushleft}
  \hspace*{\algorithmicindent} \textbf{Input }:  memory\_model, changing\_visit\_order\_number
  \end{flushleft}
  \begin{flushleft}
  \hspace*{\algorithmicindent} \textbf{Output}:DATA\_RACE, REDUNDANT\_BARRIERS
  \end{flushleft}
  \small{
\begin{algorithmic}[1]
    \Function{EXAMINE\_MEMORY\_MODEL}{}
    \State DATA\_RACE $\leftarrow$ False
    \State REDUNDANT\_BARRIERS = dict()
    \For{memory\_unit in memory\_model}
    \For{visit\_order in memory\_unit}
    \State tuples $\leftarrow$ get\_tuples\_by\_order(visit\_order)
    \For{thread performing write in tuples}
    \State other\_ts $\leftarrow$ get\_different\_threads(thread, tuples)
    \For{t in other\_ts}
    \If{in\_same\_warp(t, thread)}
    \If{using\_same\_stmt(t, thread)}
    \State DATA\_RACE $\leftarrow$ True
    \EndIf
    \If{has\_branch\_divergence(t, thread)}
    \State DATA\_RACE $\leftarrow$ True
    \EndIf
    \Else
    \State DATA\_RACE $\leftarrow$ True
    \EndIf
    \EndFor
    \EndFor
    \EndFor
    \State barrier\_dict = dict()
    \For{visit\_order in memory\_unit}
    \State next\_order $\leftarrow$ visit\_order + 1
    \State current $\leftarrow$ get\_tuples\_by\_order(visit\_order)
    \State target $\leftarrow$ get\_tuples\_by\_order(next\_order)
    \If{can\_merge\_without\_race(target, current)}
    \State barrier $\leftarrow$ get\_split\_barrier(next\_order, memory\_unit)
    \State barrier\_dict[barrier] ++
    \EndIf
    \EndFor
    \EndFor
    \For{barrier in barrier\_dict}
    \State REDUNDANT\_BARRIERS[barrier] $\leftarrow$
    \State is\_redundant(barrier\_dict[barrier], \State changing\_visit\_order\_number[barrier])
    \EndFor
    \State \textbf{return} DATA\_RACE, REDUNDANT\_BARRIERS
    \EndFunction
\end{algorithmic}
    }
    \remove{\lingming{1. rename has\_race to DATA\_RACE to be uniform with barrier divergence; 2. barrier\_lst is not available in the algorithm!; 3. rename redundant barriers to REDUNDANT\_BARRIERS}}
\end{algorithm}

\section{Experiment}

We have extensively evaluated the effectiveness and efficiency of \simulee{} in terms of detecting synchronization bugs: 
\begin{itemize} 
\item We choose all the synchronization bugs found in the 5 studied projects and apply \simulee{} to detect them. 
\item We use \simulee{} to detect new synchronization bugs of the 5 studied projects and \newproject{} additional projects according to their history (i.e., still actively maintained) and recent popularity (i.e., > 100 stars), i.e., CudaSift~\cite{cudasift} (241 stars, 103 commits, 2.4K LoC), CUDA-CNN~\cite{cudacnn} (111 stars
, 247 commits, 12K LoC), cudpp~\cite{cudpp} (202 stars, 302 commits, 58K LoC), gunrock~\cite{gunrock} (483 stars, 1467 commits, 7.8K LoC).
\item We compare \simulee{} against the open-source automatic CUDA bug detection tool \gklee{} in terms of their effectiveness and efficiency of detecting previously unknown bugs for all the projects.
\end{itemize}

\subsection{Experimental Environment and Setup}
\label{sec_setup}
We performed our evaluation on a desktop machine, with Intel(R) Xeon(R) CPU E5-4610 and 320 GB memory. The operating system is Ubuntu 16.04. For Evolutionary Programming of ``Automatic Input Generation'' in \simulee{}, the population is set to be 50, and the generation is set to be 3. Note that the \simulee{} webpage~\cite{simulee} includes more experimental results under different settings for Evolutionary Programming.
\subsection{Result Analysis}

\subsubsection{Effectiveness}
First, we applied \simulee{} to detect the synchronization bugs of the 5 studied projects. The experimental results suggest that \simulee{} can successfully detect 20 out of 27 synchronization bugs, including 12 data race bugs, 6 barrier divergence bugs, and 2 redundant barrier functions. The bugs that \simulee{} fails to detect are caused by lossy implementation logic which can be fixed only by completely refactoring the overall code structure, including abandoning the synchronization mechanisms, e.g., \commit{} ``\codeIn{b3e927edcb}'' from the project ``\codeIn{arrayfire}''.

Next, we further applied \simulee{} to detect previously unknown synchronization bugs on the total \totalproject{} projects. In addition to the five studied projects, we adopted another \newproject{} CUDA projects, i.e., CudaSift, CUDA-CNN, cudpp and gunrock. The experimental results are shown in Table \ref{tab_newbug}, where it can be observed that we successfully detected \totalnew{} bugs in total (TT), including 15 data race bugs (DR), 1 barrier divergence bug (DB), and 10 redundant barrier function bugs (RB)\remove{\lingming{8+1+10<26! Also, can we add the detailed number of datarace, redundant barrier, and barrier divergence bugs in table 3?}}. To date, 8 redundant barrier function bugs, 1 data race bug and 1 barrier divergence bug have been confirmed by the corresponding developers. 
To be specific, the developers of CudaSift and cudpp responded as follows: 
\begin{displayquote}
\textit{``Yes, there is a bit of cleaning up to do there. Sometimes when I detect oddities in the output, I add an unnecessary synchronization just in case. In fact those things should be all run on the same thread, since it cannot be parallelized anyway. Thank you for pointing it out.'' }--- CudaSift

\textit{``I think you're right...\remove{ This code, I might say, is a decade old.} There are considerably faster ways to do matrix multiply calls...''} --- cudpp
\end{displayquote}

Since barrier divergence is a undefined behavior, it may not hang on every situation. The developers of gunrock responded as follows:
\begin{displayquote}
\textit{``I do see what @Stefanlyy / @eagleShanf mean for the divergence issue, and surprise the code didn't hang.\remove{ There is a different implementation under the dev-refactor branch, and it doesn't use \_\_syncthreads(). It should work better, but I'm not 100\% sure.}''} 
\end{displayquote}

\begin{table}
\centering
\caption{\label{tab_newbug} Bugs Detected}
\begin{adjustbox}{width=\columnwidth}
\begin{tabular}{l|llll|llll|llll|llll}
\hline
\multirow{2}{*}{\bf{Projects}} & 
\multicolumn{4}{c|}{\bf{Detected}}&
\multicolumn{4}{c|}{\bf{Confirmed}} & 
\multicolumn{4}{c|}{\bf{Under Discussion}} & \multicolumn{4}{c}{\bf{Nonresponse}}\\
\cline{2-17}
& \bf{TT} & \bf{DR} & \bf{RB} &\bf{BD} & \bf{TT} & \bf{DR} & \bf{RB} &\bf{BD} & \bf{TT} &\bf{DR} &\bf{RB} &\bf{BD} &\bf{TT} &\bf{DR} &\bf{RB} &\bf{BD}\\
\hline\hline
\bf{kaldi}&11&10&1&0&1&0&1&0&6&6&0&0&4&4&0&0\\
\bf{thundersvm}&4&4&0&0&0&0&0&0&4&4&0&0&0&0&0&0\\
\bf{CudaSift}&4&0&4&0&4&0&4&0&0&0&0&0&0&0&0&0\\
\bf{CUDA-CNN}&1&0&1&0&0&0&0&0&0&0&0&0&1&0&1&0\\
\bf{cudpp}&3&0&3&0&3&0&3&0&0&0&0&0&0&0&0&0\\
\bf{gunrock}&3&1&1&1&2&1&0&1&1&0&1&0&0&0&0&0\\
\hline
\bf{Total}&26&15&10&1&10&1&8&1&11&10&1&0&5&4&1&0\\
\hline
\end{tabular}
\end{adjustbox}
\end{table}

In addition, 11 bugs, including 10 data race bugs and 1 redundant barrier function bug, are still being actively discussed by developers since some developers assumed users should acquire the prior knowledge to set the dimensions and arguments correct. We label such bugs ``under discussion'' as stated in Table \ref{tab_newbug}. 
For example, the developers of kaldi responded as follows:

\begin{displayquote}
\textit{``You could do that. Normally, though, CUDA does not expect reduction
operations to be called with more than one block. So something
like assert(gridDim.x == 1) would be better.''} 
\end{displayquote} 

In order to further evaluate the effectiveness of \simulee{}, we compare its capability in finding previously unknown bugs out of the \totalproject{} projects against \gklee{}~\cite{gklee} which is concolic-execution-based CUDA bug detector. Since \gklee{} is only designed to detect data race bugs, Table \ref{tab_gklee} shows the results for all the kernel functions from all the 9 projects that both \simulee{} and \gklee{} are applicable to and the \gklee's test-sample kernel function. In particular, the top one entry in Table \ref{tab_gklee} is \gklee{}'s test-sample kernel function, the bottom two are the kernel functions without synchronization bugs, and the rest are detected with bugs by \simulee{}\footnote{Please refer to \simulee{} webpage~\cite{simulee} for the detailed bugs and kernel functions.}. 

From Table \ref{tab_gklee}, it can be observed that \simulee{} can correctly detect all the bugs while \gklee{} can only correctly detect one bug which is actually from its own test sample. The reasons why \simulee{} performs better than \gklee{} can be inferred as follows. \simulee{} is built with a robust mechanism for bug detection, e.g., automatic setup for running environment, sufficient collection of runtime information, and simple yet complete bug detection mechanism. \gklee{}, on the other hand, suffers from severe path explosion problem such that it has to adopt pruning techniques for better efficiency while risking wrongly pruning buggy branches under some circumstances, e.g., complicated programs with loops. 
Also, \gklee{} uses the generation strategy based on traditional code coverage. In cases that code coverage cannot relate to memory access conflicts, the test cases generated by \gklee{} may not converge to a setting that can trigger synchronization bugs.
\remove{We conclude that \simulee{} is powerful in detecting synchronization bugs of CUDA programs.
}
\subsubsection{Efficiency}
To evaluate the efficiency of \simulee{}, we also compare it against \gklee{} on the 5 projects in terms of the runtime latency in detecting these bugs with the timeout set to be 6 hours.
From Table \ref{tab_gklee}, it can be observed that \simulee{} can detect all the bugs while \gklee{} can only detect one bug that is from its own testing samples and timed out on 4 bugs\remove{\lingming{we never say what is our time limit! 1 hour?}}. In other words, \gklee{} cannot detect any of the real-world bugs collected in our study. 
Specifically, for the four bugs that \gklee{} does not time out for, \simulee{} is slightly slower than \gklee{} due to Evolutionary Programming, but is still able to finish within seconds. On the other bugs, \gklee{} suffers from severe path explosion problems and takes long time before enumerating all the possible executions especially when it comes to complex programs with loops, while \simulee{} is still able to finish analysis within seconds.
From the experimental results, we can conclude that \simulee{} is a lightweight and scalable detection framework that can efficiently detect various synchronization bugs. 

\begin{table}
\centering
\caption{\gklee{} vs \simulee{}}
\label{tab_gklee}
\begin{adjustbox}{width=0.45\textwidth}
\begin{tabular}{|c||c|c|c|c|}

\hline
\bf{Kernel function} & \bf{\gklee{} time} & \bf{\simulee{} time} & \bf{\gklee{} report} & \bf{\simulee{} report} \\
\hline
\hline
1 & 343ms & 627ms& r\&w sync & r\&w sync \\
2 & 243ms & 1895ms & no sync & w\&w sync \\
3 & 486ms & 1593ms &no sync & w\&w sync \\
4 & 507ms & 1938ms &no sync & w\&w sync \\
5 & 655ms& 1472ms & no sync & w\&w sync \\
6 & Timeout & 6253ms & N/A & r\&w sync \\
7 & Timeout & 7897ms & N/A & w\&w sync \\
8 & Timeout & 2028ms & N/A & w\&w sync \\
9& Timeout & 1586ms & N/A & w\&w sync \\
10 & Timeout & 3753ms& N/A & no sync \\
11  & Timeout & 4380ms & N/A & no sync \\
\hline
\end{tabular}

\end{adjustbox}

\end{table}

\section{Threats to Validity}
\label{sec_validity}
The threats to external validity mainly lie in the subjects and faults. Though the studied projects may not represent the overall project distributions, they are selected such that the overall covered features of the CUDA projects can be maximized. On the other hand, the way that the studied bugs are derived by analyzing the commit messages may cause some false positives. In order to reduce such threat, we collect a large number of real bugs (319), which is much more than the closely related work, such as 175 in \cite{Zhang:2018:EST:3213846.3213866} and 70 in \cite{Liu:2014:CDP:2568225.2568229}. To our best knowledge, this is so far the largest study on CUDA bugs. 

The threats to construct and internal validity may lie in the different understanding towards the definition on the bug symptoms. Some symptoms might not appear to be problematic to somebody due to one's different standpoint. For instance, some developers might feel tolerant towards \perform{} issues and ignore the effectiveness of \simulee{} when it can detect them. To reduce such threat, we design two rounds of filtering to try our best to make sure that the derived bugs indeed reflect the errors that programs undergo. Next we manually check both the description and the source code to clearly understand the causes. In the end, we compare the derived symptoms against the corresponding commit messages to make sure that each bug falls into the correct category. Moreover, the feedback from the developers on our bug report submissions also appear to be supportive on our bug understandings.

\section{Related Work}
\label{sec_related}
As our work investigates the automatic bug detection techniques for CUDA programs through empirical studies, we summarize the related work into two parts: empirical studies on CUDA programs and techniques of CUDA bug detection.

\textbf{Empirical Studies}
There are several existing work that study bugs and other features on CUDA programs. For instance, Yang et al. \cite{icpp2012} delivered 
the empirical study on the features of the performance bugs on CUDA programs, 
Burtscher et al. \cite{CUDAir} studied the control-flow irregularity and memory-access irregularity and found that both irregularities are mutually dependent and exist in most of kernels. Che et al.\cite{CUDAapp} examined the effectiveness of CUDA to express with different sets of performance characteristics. Some researchers are keen on the comparisons between CUDA and OpenCL. For instance, Demidov et al. \cite{CUDAcpp} compared some C++ programs running on top of CUDA and OpenCL and found that they work equally well for problems of large size. Du et al. \cite{opencl}, on the other side, studied the discrepancies in the OpenCL and CUDA compilers' optimization that affect the associated GPU computing performance. 

\textbf{CUDA bug detection}
Several approaches that detect CUDA bugs are static/dynamic-analysis-based \cite{gklee}\cite{cudasmt}\cite{barracuda}\cite{CUDASE}\cite{PUG}\cite{GPUverify1}\cite{GPUverify2}\cite{GPUverify3}. Though they can be effective, they are also argued to be time costly \cite{curd}. A lot of the research concentrate on detecting the specific data race bugs. In addition to many aforementioned work, LDetector \cite{ld} instrumented compiler to detect races by using diffs between memory snapshots. Boyer et al. \cite{boyer2008ada} detected data race on GPU emulators instead on real GPU hardware.
Many tools have been developed to inspect CUDA programs. For instance, \gklee{}~\cite{gklee} employed
concolic execution-based verification and test-case reduction heuristics for CUDA program detections\remove{\lingming{the sentence is problematic, which test-case reduction? standalone techniques or part of gklee?}}. It was scaled as using the technique of Parameterized Flows~\cite{Parametricflow}. 

One closely-related work with \simulee{} is a test-amplification-based bug detection approach~\cite{Leung} that amplified the result of a single running test to combine it with static analysis such that the set of all
inputs and interleavings could be verified. Though the idea of injecting testing philosophy into CUDA programs is similar with \simulee{}, \simulee{} advances in (1) it is a general-purpose and fully automated bug detection framework that can detect various synchronization types while \cite{Leung} only handles data race and requires manual inputs; (2) \simulee{} only needs to run the code relevant to kernel function execution, while \cite{Leung} needs to run the whole program life cycle which leads to much larger overhead. (3) \cite{Leung} suffers from the incorrect input regarding synchronization and loss of effectiveness while \simulee{} does not have these limitations.

\section{Conclusions}
\label{sec_con}
In this paper we conduct an extensive study on CUDA program bugs. It can be concluded from the study results that the bugs occur mostly in kernel functions, where the cross-platform generic errors are the major bug symptoms. It can also be observed that the synchronization bugs can be extremely challenging to handle. Therefore, we develop a fully automated approach, namely \simulee{}, that can successfully detect synchronization bugs efficiently based on the auto-generated running environment. Specifically, \simulee{} can detect most of the synchronization bugs out of the studied projects, and even successfully detected \totalnew{} previously unknown bugs which have never been reported/detected before. In addition, \simulee{} can achieve better effectiveness and efficiency than \gklee{}.
\remove{In future, 
we hope to extend \simulee{} to detect more complicated synchronization bugs, e.g., the ones regarding inferior performance, that cannot be handled in this paper. Furthermore, we may proceed to developing new approaches that could automatically repair and synthesize CUDA program synchronization bugs. }

\newpage
\clearpage

\bibliographystyle{ACM-Reference-Format}
\bibliography{cuda-bug}

\end{document}